\documentclass[twocolumn]{article}
\usepackage{graphicx}
\begin{document}

The idea of using Josephson junctions as sources of
electromagnetic radiation is promising owing to their small
dimensions, good tunability, and capability of operating at
frequencies up to several hundred gigahertz~\cite{barone}.
However, the power of radiation available from a single junction
is not sufficient for most applications, which necessitates using
arrays of junctions. A stack arrangement of long Josephson
junctions (LJJs)~\cite{sbp} arouses interest because of possible
improvements of the properties of LJJ oscillators in terms of
impedance matching, output power, and integration level. The
low-$T_c$ superconductor thin film technology allows growth of
high-quality multilayers with many Josephson tunnel barriers (for
example, Nb/Al-${\rm AlO_x}$/Nb) stacks). Moreover, the discovery
of an intrinsic Josephson effect in some high-$T_c$
superconductors such as BSCCO convincingly showed that these
materials are essentially natural superlattices of LJJs formed on
the atomic scale~\cite{kleiner1}. Recent theoretical
investigations and experiments showed that the inductive coupling
between adjacent junctions leads to diverse and nontrivial dynamic
behaviour patterns of such
structures~\cite{jap95,japan,ustsak98,shitov96}.

One possible way to produce coherent radiation from stacks of LJJs
is to form a regular Josephson vortex lattice (JVL) and move it by
an external current. In order to produce maximum radiation from a
stack at a given lattice velocity and external magnetic field, a
rectangular arrangement of vortices, when vortices in neighboring
layers are located one over another,
 is most
preferable. Such a lattice is feasible provided the corresponding
solution is stable. In the present paper we show that in a
two-junction stack a rectangular JVL is unstable at low
velocities, and stability of such a solution can be achieved at
high velocities of JVL, provided there is a large external
magnetic field, or large enough damping, or small stack length.
Specifically, the above conditions may explain the results of
numerical~\cite{jap95,japan,ustsak98} and
experimental~\cite{shitov96} investigations in which a possibility
of existence of rectangular JVL in stacks of two and more LJJs is
shown. Our result complements the results reported in~\cite{vglen}
where the authors argue that a rectangular JVL is stable at high
velocities in an infinite stack. The incompleteness of this result
is explained by the fact that the authors of~\cite{vglen} have not
taken into account all possible perturbations of the solution in
the form of a rectangular JVL.

Let us consider the set of equations describing a simplest layered
structure --- a stack consisting of two LJJs with magnetic
coupling between the layers~\cite{sbp,vglen,bul94}:
\begin{equation}
\partial^2_x (\varphi_{1,2}+\eta\varphi_{2,1})=
(\partial^2_t + \gamma \partial_t ) \varphi_{1,2}
+\sin\varphi_{1,2} -j.\label{eqn1}
\end{equation}
Here $\varphi_{1,2}$, $\gamma$, $j$ are the Josephson phase
difference, damping constant, and bias current density,
respectively. The magnetic coupling parameter is denoted by
$\eta$. It is determined by the formula~\cite{sbp}:
\begin{equation}
\eta= \lambda (d \sinh{\frac t \lambda} +2\lambda \coth{\frac t
\lambda})^{-1}, \label{gamma}
\end{equation}
where $\lambda$ is the London penetration depth,
$t$ is the thickness of the superconducting layer, $d$ is the distance between two superconducting
layers. We start with the assumption that the system is infinite in space.

To investigate a two-junction stack it is convenient to introduce new variables
$\varphi_\pm =(\varphi_1 \pm \varphi_2)/2$,
which obey the set of equations:
\begin{equation}
c^2_{\pm}\partial^2_x \varphi_{\pm}= (\partial^2_t+ \gamma
\partial_t ) \varphi_{\pm} +\sin\varphi_{\pm} \cos \varphi_{\mp}
-j_{\pm},\label{sym}
\end{equation}
where $c^2_+ =1$, $c^2_-=(1-\eta)/(1+\eta)$, $j_+ =j$, $j_- =0$.
We have renormalized the coordinate
$x_{new}=x_{old}/\sqrt{1+\eta}$ in Eqs.~(\ref{sym}).

The set of Eqs.~(\ref{sym}) has a solution describing rectangular
JVL. Assuming the external magnetic field to be high we can write
down the analytical expressions for this solution \cite{lebst}:
\begin{equation}
\varphi_+^0 =h(x-ut) +{\rm Im} \frac{e^{ih(x-ut)}} L, \quad
\varphi_-^0 \equiv 0, \label{syms}
\end{equation}
where $L=-h^2(1-u^2)+i\gamma u h$, $u$ is the JVL velocity, $h \gg
1 $ is the dimensionless external magnetic field. Velocity $u$ and
damping $\gamma$ are related through the energy balance
condition~\cite{mcl78} which is actually the current-voltage
characteristic of the stack with the rectangular JVL:
\begin{equation}
j=-\gamma u h +\frac 1 2 {\rm Im} \frac 1 L .\label{energy}
\end{equation}
We note that the solution in the form (\ref{syms}) is valid only
provided $|L^{-1}| \ll 1$. If $h \gamma \gg 1$ the previous
condition is satisfied at all velocities otherwise it breaks in a
region near $u=1$. This region corresponds to the peak in the
current-voltage characteristic (\ref{energy}).

In order to investigate the stability of rectangular JVL we search
for the solution of~(\ref{sym}) in the form
$\varphi_{\pm}=\varphi_{\pm}^0 +\delta\varphi_{\pm}$ where
$\varphi_{\pm}^0$ are given by~(\ref{syms}) and
$\delta\varphi_{\pm}$ are small perturbations
($|\delta\varphi_{\pm}| \ll 1$). Substituting this solution
into~(\ref{sym}) and neglecting the terms nonlinear in
$\delta\varphi_{\pm}$ we obtain:
\begin{equation}
c^2_{\pm}\partial^2_x \delta\varphi_{\pm} =(\partial^2_t
+\gamma\partial_t ) \delta\varphi_{\pm} +\cos{\varphi_+^0}\cdot
\delta\varphi_{\pm} , \label{sym1}
\end{equation}
where $\cos{\varphi_+^0} \approx \cos{h(x-ut)} +{\rm Re
}((1-e^{2ih(x-ut)})/2L)$. We will refer to $\delta\varphi_+$ and
$\delta\varphi_-$ as symmetrical and antisymmetrical
perturbations, respectively. The set~(\ref{sym1}) is actually two
independent equations so we can analyze them separately. Thus we
have divided the problem of rectangular JVL stability into two
ones --- the problems of stability with respect to symmetrical and
antisymmetrical perturbations.

We start with an analysis of the "subluminal" ($u<1$) solution
stability. The problem of a rectangular JVL stability with respect
to symmetrical perturbations is similar to the problem of
stability of the periodical vortex chain in LJJ which was solved
in~\cite{lebst}. Thus a "subluminal" rectangular JVL is stable to
symmetrical perturbations. To analyze the stability
of~(\ref{syms}) with respect to antisymmetrical perturbations we
use Eq.~(\ref{sym1}) for $\delta\varphi_-$. This equation is a
relativistic invariant with $c_-$ being the characteristic
velocity of antisymmetrical perturbation. From~(\ref{gamma}) and
the expression for $c_-$ it is seen that $c_- <1$ at any stack
parameters. In other words, antisymmetrical perturbations are
always slow compared with the symmetrical ones. Therefore, to
investigate the "subluminal" solution stability it is necessary to
distinguish between two cases: $u<c_-$ and $u>c_-$.

Let us first consider the case $u< c_-$. We perform the Lorentz
transformation in Eq.~(\ref{sym1}) for $\delta\varphi_-$:
$$\xi=\frac{x-vt}{\sqrt{1-(v/{c_-})^2}},\qquad
\tau=\frac{t-(v/c^2_-)x}{\sqrt{1-(v/{c_-})^2}} $$
with velocity $v=u$. Introducing $\psi\equiv\delta\varphi_-$ we
obtain the equation
$$
c^2_- \psi_{\xi\xi} =\psi_{\tau\tau} +\frac \gamma
{\sqrt{1-(u/c_-)^2}}(\psi_\tau - u \psi_\xi)
+\cos{\varphi_+^0}\cdot \psi
$$
where the parameter
depends only on the coordinate $\xi$. After the renormalization of
the coordinate $h\sqrt{1-(u/c_-)^2} \xi_{old} =2\xi_{new}$  and
time $h\sqrt{1-(u/c_-)^2} \tau_{old} =2\tau_{new}$ and
introduction of the small parameter $\mu=4h^{-2}(c_-^2-u^2)^{-1}$
we have:
\begin{equation}
\psi_{\xi\xi} + \Gamma_\xi \psi_\xi = \frac 1
{c_-^2}(\psi_{\tau\tau} +\Gamma_\tau \psi_\tau )
+\mu[\cos{2\xi}+{\rm Re} \frac {1-e^{4i\xi}} {2L}]\psi,
\label{asym3}
\end{equation}
where $\Gamma_\xi=\mu hu\gamma/2$, $\Gamma_\tau=\mu h\gamma
c_-^2/2$. We look for the solution of~(\ref{asym3}) in the form of
the Fourier integral $\psi(\xi,\tau)= \int_{-\infty}^\infty
\tilde\psi(\xi,\omega) e^{-i\omega \tau} \frac{d\omega}{2\pi}$.
The equation for the Fourier image of $\psi (\xi,\tau)$ is
\begin{equation}
\tilde\psi_{\xi\xi}+\Gamma_\xi \tilde\psi_\xi=-\frac{\omega^2
+i\omega\Gamma_\tau }{c_-^2} \tilde\psi+ \mu[\cos{2\xi} +{\rm Re}
\frac {1-e^{4i\xi}} {2L} ]\tilde\psi. \label{hill1}
\end{equation}
\begin{figure}
\includegraphics[width=8.6cm]{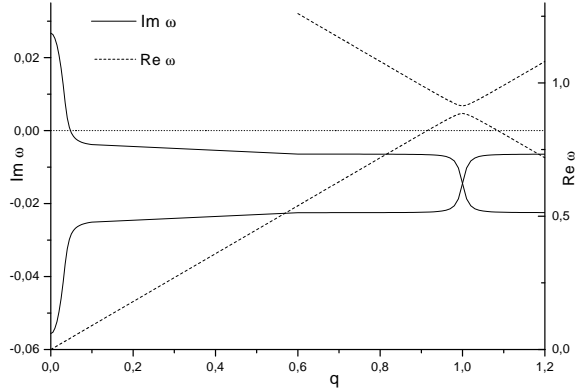}
\caption{Eigenfrequency $\omega(q)$ spectrum as a function of
quasimomentum $q$ for antisymmetric perturbations (case $u <
c_-$)} \label{disp1}
\end{figure}
According to the Bloch theorem, the solutions of~(\ref{hill1})
have the form $\tilde\psi (\xi)=\exp {(iq\xi)} U_q (\xi)$, where
$q$ is the quasimomentum and $U_q(\xi)$ is the function with the
period $\pi$. Let us find the eigenfrequency spectrum $\omega(q)$.
At $\mu = 0$ the spectrum is
\begin{equation}
\omega (\omega+i\Gamma_\tau) =c_-^2 q(q-i\Gamma_\xi).
\label{unpert}
\end{equation}
At $\mu \neq 0$ the maximum perturbations of the spectrum are
achieved near the middle point ($q=0$) and the edges ($q=\pm 1$)
of the first Brillouin zone. In the vicinity of $q=0$ we search
for the solution of~(\ref{hill1}) in the form
$$
\tilde\psi(\xi)=e^{iq\xi}[a_0 + a_2 e^{2 i\xi}+a_{-2}e^{-2 i\xi}],
$$
where $a_0, a_{\pm 2}$ are constants. After substituting it
into~(\ref{hill1}) we obtain the dispersion characteristic as the
condition for $a_{0, \pm 2}$ at which the solution
of~(\ref{hill1}) is not equal to zero:
\begin{equation}
-(q-\frac{i\Gamma_\xi} 2)^2 +c_-^{-2} (\omega+\frac{i\Gamma_\tau}
2 )^2 -\mu\alpha =-\frac {\mu^2} 8, \label{dd11}
\end{equation}
where $\alpha={\rm Re}\, (2L)^{-1} -\gamma^2 /4$. Near $q=1$ the
solution has the form
$$
\tilde\psi(\xi)=e^{i(q-1)\xi}[a_1 e^{i\xi}+a_{-1}e^{-i\xi}],
$$
where $a_{\pm 1}$ are constants. Substitution of this expression
into~(\ref{hill1}) gives
\begin{equation}
(\frac{\omega+i\Gamma_\tau /2} {c_-} -\frac{\mu\alpha} 2 -1)^2 -
(q-\frac{i\Gamma_\xi} 2 -1)^2 = \frac {\mu^2} {16}.
\end{equation}
Far from the middlepoint and the edges of the first Brillouin zone
the spectrum remains unperturbed and is given by the
formula~(\ref{unpert}).

The dependencies of real and imaginary parts of eigenfrequency
$\omega(q)$ are shown in Fig.{\ref{disp1}}. It is seen that at
small $q$ some roots of the dispersion equation (\ref{dd11}) have
positive imaginary parts. This means that perturbations with small
$q$ will exponentially grow with time, i.e. the solution
(\ref{syms}) is unstable. We see that the instability which is
obvious in the case of low density chains (due to repulsion of
vortices in the neighbouring layers) is not changed by stability
in the case of denser chains. This result is in agreement with the
one obtained in \cite{vglen} and reflects the fact that the vortex
chains in the neighbouring layers tend to shift and form the
triangular JVL.

Let us now consider the case of high velocities $c_- < u < 1$. As
before, we perform the Lorentz transformation in Eq.~(\ref{sym1})
for $\delta\varphi_-$ but now with the velocity $v=c^2_- /u$.
Introducing $\psi \equiv \delta\varphi_-$ we obtain the equation
$$
c^2_- \psi_{\xi\xi} =\psi_{\tau\tau} +\frac \gamma
{\sqrt{1-(c_-/u)^2}} (\psi_\tau-\frac {c_-^2} u \psi_\xi)
+\cos{\varphi_0^+}\cdot \psi,
$$
where the parameter depends only on the time $\tau$. This equation
turns into
\begin{equation}
\psi_{\tau\tau}+\Gamma_\tau \psi_\tau =c_-^2 (\psi_{\xi\xi}
+\Gamma_\xi \psi_\xi) -\mu(\cos{2\tau}+{\rm
Re}\frac{1-e^{4i\tau}}{2L})\psi, \label{asym4}
\end{equation}
where $h\sqrt{u^2-c_-^2}\tau_{old}=2\tau_{new}$,
$h\sqrt{u^2-c_-^2}\xi_{old}=2\xi_{new}$, $\mu
=4h^{-2}(u^2-c_-^2)^{-1}$, $\Gamma_\tau =\mu u h \gamma /2$,
$\Gamma_\xi =\mu h \gamma /2$. We look for the solution
of~(\ref{asym4}) in the form of the Fourier integral
$\psi(\xi,\tau)=\int_{-\infty}^\infty \tilde\psi(k,\tau)
e^{-ik\xi} \frac{dk}{2\pi}$. The equation for the Fourier image of
$\psi(\xi,\tau)$ is
\begin{equation}
\tilde\psi_{\tau\tau}+\Gamma_\tau \tilde\psi_\tau = -c_-^2 (k^2
-ik\Gamma_\xi)\tilde\psi -\mu(\cos{2\tau}+{\rm
Re}\frac{1-e^{4i\tau}}{2L}) \tilde\psi. \label{mat2}
\end{equation}
The solution of~(\ref{mat2}) has the form
$\tilde\psi(\tau)=\exp{(-i\varepsilon \tau)}U_{\varepsilon}(\tau)$
where $\varepsilon$ is the quasienergy and $U_{\varepsilon}(\tau)$
is the function with the period $\pi$. Let us find the spectrum
$\varepsilon (k)$ by the scheme used above. In the vicinity of
$\varepsilon=0$ the spectrum is
\begin{equation}
-(\varepsilon +\frac{i\Gamma_\tau} 2 )^2 +c_-^2
(k-\frac{i\Gamma_\xi} 2 )^2 +\mu\alpha = -\frac {\mu^2} 8,
\end{equation}
and near $\varepsilon=1$:
\begin{equation}
(c_-(k-\frac{i\Gamma_\xi} 2 )+\frac{\mu\alpha} 2 -1)^2 -
(\varepsilon +\frac{i\Gamma_\tau} 2 -1)^2 =\frac {\mu^2} {16}.
\label{dd22}
\end{equation}
\begin{figure}
\includegraphics[width=8.6cm]{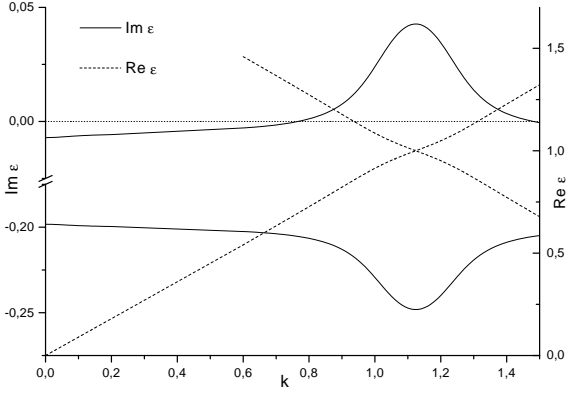}
\caption{Quasienergy $\varepsilon(k)$ as a function of wavenumber
$k$ for antisymmetric perturbations (case $c_- < u < 1$)}
\label{disp2}
\end{figure}
It follows from the Eq.~(\ref{dd22}) that at certain $k$ the roots
of the dispersion equation may have positive imaginary part as it
is shown in Fig.\ref{disp2}. ${\rm Im}\, \varepsilon(k)$ reaches
its maximum value at $k = c_-^{-1}$. At small $\gamma$ $({\rm
Im}\, \varepsilon(k))_{max} \approx \mu/4 -\Gamma_\tau /2$.
Therefore, the perturbation with $k = c_-^{-1}$ depends on time
like
$$
\tilde\psi(k,\tau) \sim \exp{[(\frac {\mu}4 -\frac{\Gamma_\tau}2
)\tau_{new}}] = \exp{[\frac{(h^{-1}-u\gamma)\tau_{old}}{2\sqrt{u^2
-c_-^2 }}]}.
$$
It is seen from this expression that the solution (\ref{syms}) is
either stable or unstable depending on sign of the difference
$h^{-1}-u\gamma$. If $h u\gamma <1$ the solution (\ref{syms}) is
parametrically unstable. The region of $k$ corresponding to the
growing perturbations is equal to
\begin{equation}
\Delta k \approx \frac 2 {c_-} \sqrt{\frac{\mu^2}{16}-\frac
{\Gamma_\tau ^2} 4 }. \label{kinst}
\end{equation}
This parametric instability may be "suppressed" either by
increasing the external magnetic field $h$ or by increasing the
damping $\gamma$. We would like to emphasize that the instability
appears due to the periodicity of the solution. Therefore the
results obtained in~\cite{ge93} for the isolated vortices cannot
be applied to the case of periodic vortex chains.

It remains to investigate the stability of the "superluminal" JVL
($u>1$). We start with the analysis of the stability with respect
to symmetrical perturbations. Substituting the
solution~(\ref{syms}) into Eq.~(\ref{sym1}) for $\delta\varphi_+$
and applying the Lorentz transformation
$$
\xi=\frac{x-vt}{\sqrt{1-v^2}},\qquad
\tau=\frac{t-vx}{\sqrt{1-v^2}}
$$
with $v=u^{-1}$,
we obtain
\begin{equation}
\psi_{\xi\xi} =\psi_{\tau\tau} +\frac \gamma {\sqrt{1-(1/u)^2}}
(\psi_\tau-\frac 1 u \psi_\xi) +\cos{\varphi_0^+}\cdot \psi,
\end{equation}
where $\psi \equiv \delta\varphi_+$. In this
equation the coefficient at $\psi$ depends only on time $\tau$ and
is periodical. Analyzing this equation by the method used for the
case $c_- <u<~1$ we arrive at the conclusion that there are
shortwave perturbations which depend exponentially on time $\tau$
and the solution (\ref{syms}) may be stable or unstable depending
on sign of the difference $h^{-1} -u\gamma$. The same result is
obtained also for the antisymmetric perturbations in the case $u
>1$. The main feature of this result is that the stability
condition $h^{-1}< u\gamma$ may be achieved by increase of $u$.
However, the alternate component of the electric field $\varphi_t$
of such a solution has the order $h^{-1} u (u^2 -1)^{-1}$ and is
small. Therefore, the power of radiation from the edge of the
stack in this case is not sufficient for applications in spite of
the fact that the JVL is rectangular.
\begin{figure}[t]
\includegraphics[width=8.6cm]{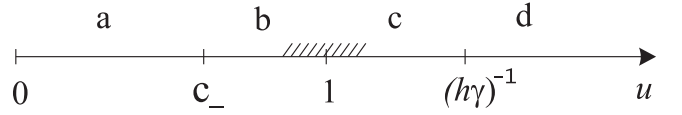}
\caption{Stability diagram for the rectangular vortex lattice
(case $h\gamma <1$). Latin letters mark the regions of different
behaviour of small perturbations: a -- longwave instability, b --
shortwave instability to antisymmetric perturbations, c --
shortwave instability to both symmetric and antisymmetric
perturbations, d -- stability. Hatching shows the region where the
solution~(\ref{syms}) is not valid.} \label{diagram}
\end{figure}

Combining the results obtained above we build the diagram
(Fig.~\ref{diagram}) showing the regions of stability and
instability in terms of the lattice velocity $u$. At $0 < u < c_-$
the solution (\ref{syms}) is unstable with respect to longwave
perturbations and formation of the triangular JVL. At $c_- < u <
(h\gamma)^{-1}$ the solution is unstable due to the parametric
resonance and the instability growth rate is proportional to
$h^{-1} -u \gamma$. In the region near $u=1$ where the condition
$|L^{-1}| \ll 1$ breaks the solution~(\ref{syms}) is not valid and
more thorough investigation is required.  Then, at $u
> (h\gamma)^{-1}$ the parametric resonance is "suppressed" and the
solution becomes stable. The diagram Fig.~\ref{diagram} is built
for the case $h\gamma <1$. When $h\gamma > 1$ the instability
region is $0 < u < (h\gamma)^{-1}$. At last, when $(h\gamma)^{-1}
< c_-$ the instability region is $0 < u < c_-$.

Consider now the case when the system is not infinite in space.
The simplest way to change to a finite system is to set periodic
boundary conditions
$$
\varphi_{1,2}(x=l,t) = \varphi_{1,2}(x=0,t) +2\pi N,
$$
where $N$ is the number of vortices trapped in each junction of
the stack, $l$ is the length of the system. The boundary
conditions for perturbations are written as below:
\begin{equation}
\delta\varphi_{\pm} (\xi_{new}=hl/2 , \tau) = \delta\varphi_{\pm}
(\xi_{new}=0, \tau).\label{bound}
\end{equation}
At $u<c_-$ the conditions~(\ref{bound}) lead to discreteness of
quasimomentum $q$ with the step $4\pi/hl$. As the instability
interval is located near $q=0$, mode with $q=0$ will always be in
this interval, i.e. it will grow with time. Hence, the periodic
boundary conditions cannot provide stability of a rectangular JVL
at small velocities. At $c_- < u <1$  the wavenumber $k$ is
discrete with the step $4\pi/hl$. At large $l$ when $\Delta k >
4\pi /hl$ at least one of the possible values of $k$ will be in
the interval of instability~(\ref{kinst}), therefore, the solution
will be unstable at $h^{-1} > u\gamma$. The necessary condition of
stability is
$$
l >\pi h c_- (c_-^2 - u^2).
$$
This means that to provide a rectangular JVL stability at $c_- < u
<1$ it is necessary to decrease the system length $l$  so that no
permitted value of $k$ is in the interval~(\ref{kinst}). At $1 < u
< (h\gamma)^{-1}$ the situation is more complicated because the
instability intervals in $k$ have different widths and are
differently located for symmetrical and antisymmetrical
perturbations. But it is clear that the decrease of $l$ lowers the
number of permitted $k$ in the instability intervals and may lead
to stability.

Let us summarize the obtained results. At low velocities a
rectangular JVL in an infinite two-junction stack is unstable with
respect to perturbations with scales much larger than the lattice
period. The physical meaning of this is that the vortex chains in
the first and second junctions tend to shift with respect to each
other to compose a triangular JVL. This is in agreement with the
result of~\cite{vglen}. At high but "subluminal" velocities the
lattice may be unstable with respect to perturbations with scales
comparable to the period of the JVL. The growth rate of this
parametric instability is determined by the difference
$h^{-1}-u\gamma$ and, therefore, the instability may be
"suppressed" by high enough external magnetic field, or by the
damping in the system, or by the increase of the lattice velocity.
 The existence of this instability complements the results obtained
 in~\cite{vglen} where the authors did not take into account
the short-wave perturbations. In the present paper we also show
that the stability of rectangular JVL at low damping is possible
in a finite two-junction stack. Thus we argue that the formation
of a rectangular JVL which is reported
in~\cite{jap95,japan,ustsak98,shitov96} is associated either with
the finiteness of the system or with the suppression of the
instability by three ways mentioned above.

This work was supported by the Russian Foundation for Fundamental Research (Grant No.
00-02-16528). The authors are grateful to N. Gress for assistance in the preparation
of the English version of the manuscript.

\end{document}